\documentclass[a4paper, 11pt]{article}
\usepackage[english]{babel}
\usepackage{setspace}
\usepackage{graphicx}
\usepackage{float}
\usepackage{enumerate}
\usepackage{url}
\usepackage{amsmath}
\usepackage[super,numbers,sort&compress]{natbib}
\usepackage{hyperref}
\usepackage{cleveref}
\usepackage[T1]{fontenc}
\usepackage{verbatim}
\usepackage{braket}
\usepackage[affil-it]{authblk}
\usepackage[textheight=23cm,textwidth=17cm]{geometry}
\usepackage{subcaption}

\author{Moritz Th\"urlemann}
\author{Lennard B\"oselt}
\author{Sereina Riniker\thanks{Corresponding author: email: sriniker@ethz.ch, ORCID: 0000-0003-1893-4031}}
\affil{Laboratory of Physical Chemistry, ETH Zurich, Vladimir-Prelog-Weg 2, 8093 Zurich, Switzerland}

\title{\textbf{Learning Atomic Multipoles: Prediction of the Electrostatic Potential with Equivariant Graph Neural Networks}}

\date{}

\begin{document}

\maketitle

\begin{abstract}
The accurate description of electrostatic interactions remains a challenging problem for fitted potential-energy functions.
The commonly used fixed partial-charge approximation fails to reproduce the electrostatic potential at short range due to its insensitivity to conformational changes and anisotropic effects.
At the same time, possibly more accurate machine-learned (ML) potentials struggle with the long-range behaviour due to their inherent locality ansatz.
Employing a multipole expansion offers in principle an exact treatment of the electrostatic potential such that the long-range and short-range electrostatic interaction can be treated simultaneously with high accuracy. However, such an expansion requires the calculation of the electron density using computationally expensive quantum-mechanical (QM) methods.
Here, we introduce an equivariant graph neural network (GNN) to address this issue. The proposed model predicts atomic multipoles up to the quadrupole, circumventing the need of expensive QM computations.
By using an equivariant architecture, the model enforces the correct symmetry by design without relying on local reference frames. The GNN reproduces the electrostatic potential of various systems with high fidelity.
Possible use cases for such an approach include the separate treatment of long-range interactions in ML potentials, the analysis of electrostatic potential surfaces, and for the static multipoles in polarizable force fields.
\end{abstract}

\section{Introduction}
The accurate treatment of the electrostatic potential (ESP) is crucial for the description of molecular interactions \cite{Stone}. 
In classical biomolecular force fields (FFs), the electrostatic interaction is approximated using fixed, i.e. time- and conformation-invariant, partial charges \cite{FixedChargeFF}. 
Higher-order multipoles are not treated explicitly, thus neglecting the anisotropic part of the electrostatic interaction.
Short-range interactions are therefore not accurately described for most systems of biological and pharmaceutical interest \cite{ChargeAnisotropy, CrystalFF2}.
The desire for a more accurate description of the ESP motivated the development of polarizable FFs such as AMOEBA \cite{AMOEBA}, which describes the electrostatic interaction with higher moments up to quadrupoles, as well as several FFs used in crystal structure prediction \cite{CrystalFF1, PolarizableFF}.

In recent years, there has been considerable effort to replace expensive \textit{ab initio} and DFT calculations with machine-learned (ML) potentials \cite{PhysNet, DIME, SOAP, HDNNP, GDML, MachineLearningFF, SCHNET, FCHL, GilmerQuantumGNN, ANI}. In many cases, these models moved away from the historically established decomposition of the system energy into separate interactions. Instead, the total potential energy is decomposed into atomic contributions. While generally providing promising results at short range, these ML models often fail to model long-range interactions with sufficient accuracy \cite{MrBoeselt}.
In principle, the treatment of the ESP based on multipoles is able to resolve not only the long-range deficiency of ML models, but also the inaccuracy of the partial-charge approximation in fixed-charge FFs. 
However, an explicit treatment of electrons is required to obtain multipoles, which is often cost prohibitive. Predicting multipoles with ML could remove this hurdle.

Previous efforts to use ML for the prediction of electrostatic interactions have mostly focused on partial charges, with models that either solely predict partial charges \cite{Patrick, ElectronMessagePassing, MLChargeAssignment}, models that combine a local treatment with an explicit treatment of long-range interactions based on partial charges \cite{PhysNet, Behler1, Behler2}, or models that predict partial charges as an auxiliary variable for properties such as the molecular dipole \cite{MolecularDipole, DipolesCeriotti}.
Due to the smaller influence of higher-order atomic multipoles on the ESP and their orientational dependence, these terms have found much less attention. 
Particularly, the orientational dependence poses a challenge for commonly deployed invariant descriptors. 
In previous work with ML models and in several polarizable FFs, this problem was circumvented by mapping multipole components onto a local reference frame spanned by neighbouring atoms\cite{MultipoleML1, MultipoleML2, LocalFrame}. However, local reference frames may lack transferability.
In an alternative approach, Glick \textit{et al.}\cite{Multipolesherrill} proposed Cartesian message passing neural networks by introducing vectors between neighbouring atoms as features. The proposed model was used to successfully predict higher-order multipoles. However, the proposed model architecture does not enforce the proper symmetry by design. Equivariance is hence only approximately achieved through data augmentation.

Recently, ML methods that respect such symmetries were introduced \cite{TensorFieldNetworks, SE3Transformers, GroupEquivariance, EquivarianceLie, EquivariantGNN, Cormorant, GroupEquivariantCNN}.
While not in all cases used to predict equivariant properties, several models applied to chemical problems showed that introducing directional information through equi- or covariant layers improves accuracy while often requiring less parameters and training data than comparable models that do not take advantage of directional information \cite{DIME, EquivariantConvolutions, EffectEquivariance}. Sch\"utt \textit{et al.} \cite{PAINN} introduced a model based on equivariant message passing, which includes directional information and is used to predict equivariant properties. For instance, their work showed that, similarly to Ref.~\citenum{DipolesCeriotti}, the introduction of atomic dipoles resulted in an improvement when predicting the infrared and Raman spectrum of ethanol and aspirin in vacuum.

In this work, an ML model for predicting multipoles based on a standard message passing graph neural network (GNN) architecture is introduced \cite{GNN, InteractionNetwork}. GNNs can be considered a natural choice for problems in (computational) chemistry, where nodes represent atoms and edges describe bonds or interactions between atoms.
In this setting, GNNs have been applied to a wide range of problems, often delivering state-of-the-art results \cite{DIME, MolecularFingerprint, GilmerQuantumGNN, GNNCrystal}. 
The graph construction as well as the manner in which information flows through the model are the main features discriminating between different GNN flavors. Due to its wide applicability, a standard message passing architecture was chosen as introduced by Battaglia \textit{et al.} \cite{GNN, InteractionNetwork}.
Most importantly, we employ a model based on the concepts of E(n)-equivariant GNN introduced by Satorras \textit{et al.}\cite{EquivariantGNN}. 
Their proposed method provides an exceedingly simple and elegant approach, which introduces equivariance for standard message passing GNNs without adding significant model-complexity or computational overhead. While only applied to the prediction of positions and velocities in the original work, we show that this approach can be extended for the prediction of atomic multipoles. Due to its structure, the developed ML model enforces the proper symmetry by design, i.e. equivariance with respect to rotations.

\section{Theory}
\subsection{Electrostatic Potential (ESP) and Multipoles}
The ESP of a charge distribution $\rho (r')$ at a point $r$ is defined as,
\begin{equation}
    V_{\text{ES}}(r) = \int_{V'}\frac{\rho (r')}{|r-r'|}dV' .
\end{equation}
Taylor expansion of $V_{\text{ES}}(r)$ around the center of the charge distribution gives rise to the well known multipole expansion \cite{Stone},
\begin{equation}
    V_{\text{ES}}(r) = \frac{M^{(0)}}{|r|} + \frac{M_\alpha^{(1)}r_\alpha}{|r|^3} + \frac{M^{(2)}_{\alpha\beta}(3r_\alpha r_\beta - r^2\delta_{\alpha\beta})}{2|r|^5} + ... ,
\end{equation}
where $M^{(0)}$, $M^{(1)}$ and $M^{(2)}$ correspond to the monopole ($q$), the dipole ($\mu$), and the quadrupole ($\theta$), respectively. Greek indices run over the components of the coordinate system. Note that the prefactor $(4\pi\epsilon_0)^{-1}$ is omitted throughout this work for simplicity.

Following the same argumentation and assuming that a molecular charge density can be decomposed into atomic charge densities,
\begin{equation}
    \rho(r)=\sum_i \rho_i(r) ,
\end{equation}
the same concept can be applied to each atomic density, giving rise to atomic or distributed multipoles \cite{Stone, MultipoleMethod}.

\subsection{Atomic Multipoles}
Several methods for the calculation of atomic multipoles have been introduced, including distributed multipole analysis \cite{DMA}, transferable atom equivalents method \cite{TAE}, and Gaussian multipole models \cite{GMM} to name a few. In addition, several approaches that fit multipoles directly to the ESP have been proposed \cite{ESPFitting, LocalFrame}.
In a recent work, Verstraelen \textit{et al.} \cite{IS1, IS2, MBIS} proposed the minimal basis iterative Stockholder (MBIS) method. MBIS was chosen for its low computational cost and fast convergence with respect to the multipole order.
MBIS can be considered an atoms-in-molecule method, which attempts to partition the total electron density into atomic contributions. The method is based on the minimization of the Kullback-Leibler divergence between a pro-density $\rho_A^0(r)$ based on a minimal expansion in atom-centered s-type Slater functions and a target molecular density $\rho_A(r)$ \cite{Kullback1951}. 
\begin{equation}
    \rho_A^0(r) = \sum_{i=1}^{m_A}\rho^0_{Ai}(r)
\end{equation}
where $\rho_A^0(r)$ is the pro-atomic density constructed from Slater functions, i.e.
\begin{equation}
    \rho^0_{Ai}(r) = N_{Ai}f_{Ai}(r)=\frac{N_{Ai}}{\sigma_{Ai}^38\pi}\exp\bigg(-\frac{|r-R_A|}{\sigma_{Ai}}\bigg) ,
\end{equation}
with a fitting parameter $N_{Ai}$ for the electron population,  $\sigma_{Ai}$ for the width and $R_A$ as the atomic center.
The divergence between the pro-density and the target density is minimized by iteratively adjusting the population and width of each atomic shell, optimizing the KL divergence for every atom
\begin{equation}
    \Delta S[\{\rho_A\}; \{\rho_A^0\}]=\sum_{A=1}^{N_{atoms}}\int \rho_A(r)\log\bigg(\frac{\rho_A(r)}{\rho_A^0(r)}\bigg)dr.
\end{equation}
The converged pro-density can then be processed to obtain desired properties, for instance partial charges or atomic multipoles \cite{MBIS}.
 
\subsection{Graph Neural Networks}
In the following sections, we follow the notation used by Refs.~\citenum{EquivariantGNN, GilmerQuantumGNN}.
GNNs are ANN-parametrized ML models, which process graph-structured data. In their commonly used form, node, edge and/or global features are iteratively refined based on the current features. GNN models differ mainly by the features used, the way the underlying graph is constructed, as well as the updating or feature-refinement process applied \cite{GNN, GilmerQuantumGNN}.

Considering a graph $G=(V, E)$ with nodes $v_i \in V$ and edges $e_{ij} \in E$, message passing can be defined as
\begin{equation}
    \begin{aligned}
        m_{ij} &= \phi_e (h_i^l, h_j^l, a_{ij})\\
        m_i &= \sum_{j\in N(i)}m_{ij}\\
        h_i^{l+1} &= \phi_h(h_i^l, m_i) ,
    \end{aligned}
\end{equation}
where $h_i^l\in \mathbf{R}^n$ describes the hidden-feature vector of node $v_i$ after $l$ steps, $a_{ij}\in \mathbf{R}^n$ the edge feature of edge $e_{ij}$ between node $i$ and $j$, $N(i)$ denotes the set of neighbours of $v_i$, and $\phi_e$ and $\phi_h$ describe update functions, which are commonly parametrized by an artificial NN. The superscript $l$ is the current layer, or in its recurrent realization the current iteration. In the present work, atoms are represented by nodes $v$ and interactions between atoms as edges $e$.

\subsection{Equivariance}\label{sec:equivariance}
Following the formalism used in Ref.~\citenum{EquivariantGNN}, a function $\phi: X \rightarrow Y$ mapping from an input space $X$ to an output space $Y$ is called equivariant to a group action $g\in G$ if
\begin{equation}
    \phi(\hat{T}_g(x))=\hat{S}_g(\phi(x))
\end{equation}
for a transformation $\hat{T}_g$ on $X$, an equivalent transformation $\hat{S}_g: Y \rightarrow Y$ on the output space $Y$ and a group $G$ \cite{EquivariantGNN}. In other words, equivariance describes maps that commute under the action of a group \cite{GroupEquivariantCNN, GroupTheoryBook}.

For the practical case of a dipole $\mu(X)$ of an atom or molecule positioned at $X$, we require that $\hat{R}(\mu(X)) = \mu(\hat{R}(X))$ for an arbitrary rotation or reflection $\hat{R}\in\mathbf{O}(3)$. In this specific example, $\mathbf{O}(3)$ would be the group $G$, an arbitrary rotation or reflection would be the transformation $\hat{T}_g$ applied to the dipole $\mu(X)$ or the coordinates $X$, and $\mu$ would be the equivariant function $\phi$.
In other words, the dipole should rotate in the same way as the coordinates rotate. Any function computing the dipole of a system $X$ must fulfill this condition. Note that it is not possible to fulfill this condition with ML models based on commonly employed invariant descriptors.
We therefore introduce equivariance for GNNs to predict atomic multipoles following the ideas reported by Satorras \textit{et al.} in Ref.~\citenum{EquivariantGNN}.
The authors implemented $E(n)$ equivariance by introducing an auxiliary variable $x\in \mathbf{R}^n$ for each node in the graph. In their work, $x$ represents a position or velocity in Cartesian coordinates.
Further, in order to maintain equivariance, the following message passing function was proposed,
\begin{equation}
    \begin{aligned}
        m_{ij} &= \phi_e (h_i^l, h_j^l, ||x_i^l - x_j^l||^2, a_{ij}) \\
        x_i^{l+1} &= x_i^l + C\sum_{j\neq i}(x_i^l - x_j^l)\phi_x(m_{ij}) \\
        m_i &= \sum_{j\in N(i)}m_{ij} \\
        h_i^{l+1} &= \phi_h(h_i^l, m_i),
    \end{aligned}
\end{equation}
where $C$ is a normalization factor and $x_i$ denotes the equivariant vector of atom $i$.
This rule effectively separates equivariant from invariant features. In the subsequent sections we adapt this concept to the prediction of atomic multipoles of molecular systems.

\subsection{Equivariance for Multipoles}
The notation for multipoles follows the work by Burnham \textit{et al.} \cite{MultipoleMethod}.
The present approach relies on the assumption that every multipole can be decomposed into a linear combination of coefficients depending on the local environment of an atom (invariant features) and a spatial component (equivariant feature) depending on the relative orientation between atoms.
Analogous to the model described in the previous section, the coefficients are predicted based on invariant features while the spatial (i.e. equivariant) part is taken as the $k$-times outer product of $R_{ij}$,
\begin{equation}
    R_{ij}^{(k)} = R_{ij}\otimes R_{ij} \otimes ... ,
\end{equation}
where $k$ denotes the order of the multipole and $R_{ij}$ labels the vector $R_{ij}=R_j - R_i$ between two atoms. Accordingly, the following composition for $M_i^{(k)}$, the $k$-$th$ multipole of atom $i$, is proposed,
\begin{equation}\label{eq:general_multipole_rule}
    M_i^{(k)} = \sum_{j\in N(i)} c_{ij}^{(k)}R_{ij}^{(k)}
\end{equation}
with a scalar coefficient $c^{(k)}_{ij}\in\mathbf{R}$ depending on the atom pair $ij$ and $R_{ij}^{(k)}$ as the $k$-$th$ outer product of the vector pointing from atom $i$ to atom $j$. Generally, $c_{ij}^{(k)}$ is predicted based on the features of atom $i$ and atom $j$ as,
\begin{equation}
    c_{ij}^{(k)}=\phi_{ M_i^{(k)}}(h_i^n, h_j^n, a_{ij}),
\end{equation}
where $\phi_{M^{(k)}}$ denotes an ANN parametrized function, mapping two atomic feature vectors $h^n_i$ and a distance-based edge feature $a_{ij}$ to a scalar coefficient.

Initially, atomic features are iteratively updated using a standard message passing GNN. Since nodes and edges depend only on atom types and distances, these features contain purely rotation- and translation-invariant information. 
After $n$ message passing iterations, the atomic features $h_i^n$ serve as the descriptor for the prediction of the scalar coefficients in the subsequent steps. We note that $h_i^n$ could be replaced by any other invariant feature proposed in the literature. In particular, even atomic features which are purely molecular graph-based can be used while still retaining some degree of conformational dependence for atomic dipoles and quadrupoles.

Beginning with the atomic monopole, $M^{(0)}_i$ can be predicted as
\begin{equation}
    M^{(0)}_i = \sum_{j\in N(i)}R_{ij}^{(0)}\phi_{M^{(0)}} (h_i^n, h_j^n, a_{ij}) ,
\end{equation}
where $R_{ij}^{(0)}$ is the identity. As proposed by Metcalf \textit{et al.} \cite{ElectronMessagePassing}, charge conservation can be enforced elegantly by imposing anti-symmetry on $\phi_{M^{(0)}}$ with respect to the hidden features $h_i^n$ and $h_j^n$.
Alternatively, an atom-based scheme can be applied where each atomic monopole $M^{(0)}_i$ is predicted based on the hidden feature $h_i^n$ of the respective atom, i.e.
\begin{equation}\label{eq:AtomicMonopole}
    M^{(0)}_i = \phi_{M^{(0)}} (h_i^n).
\end{equation}
In this case, charge conservation can be enforced by subtracting the mean total charge of a given system ($Z$) subtracted by the mean predicted charge $\langle q \rangle = \frac{Z}{N} - \frac{1}{N}\sum_i^Nq_i$ from each atomic monopole.
In preliminary tests, the atom-based approach of Eq.~(\ref{eq:AtomicMonopole}) resulted in a better performance and is thus used in the final model.

In congruence with the equivariant updating rule, higher-order multipoles, beginning with the atomic dipole $M^{(1)}_i$, are predicted as,
\begin{equation}
    M^{(1)}_i = \sum_{j\in N(i)} R_{ij}^{(1)} \phi_{M^{(1)}} (h_i^n, h_j^n, a_{ij}).
\end{equation}
For the atomic quadrupole $M^{(2)}_i$, which is the highest-order multipole considered in this work, the following rule is used,
\begin{equation}
    M^{(2)}_i = \sum_{j\in N(i)} R_{ij}^{(2)} \phi_{M^{(2)}} (h_i^n, h_j^n, a_{ij}).
\end{equation}
If $M^{(2)}_i$ is taken as the traceless quadrupole, $R_{ij}^{(2)}$ should be detraced in the same manner. In the present work, the detraced quadrupole $M^{(2)}_i$ and outer product $R_{ij}^{(2)}$ are defined as,
\begin{equation}
    \widehat{D}A^{(2)}_{\alpha\beta}=A_{\alpha\beta} - \frac{1}{3}\sum_\chi A_{\chi\chi}\delta_{\alpha\beta} ,
\end{equation}
following the notation in Ref.~\citenum{MultipoleMethod}. Greek indices run over the three Cartesian dimensions, $\widehat{D}$ refers to the detracing operator and $A$ stands for a quadrupole $M^{(2)}_i$ or outer product of a vector with itself $R_{ij}^{(2)}$.

\subsection{ESP Derived from Multipoles}\label{sec:esp_from_multipoles}
Using the same notation, the ESP $V_{\text{ES}}$ at a point $R_j$ arising from point multipoles at $R_i$ is given as,
\begin{equation}
     V_{\text{ES}}(R_{ij})=\sum_{n=0}^\infty \frac{(2n-1)!!}{n!|R_{ij}| ^{2n + 1}}\langle M_i^{(n)}, R_{ij}^{(n)} \rangle ,
\end{equation}\label{eq:multipole_esp}
where both $M_i^{(n)}$ and $R_{ij}^{(n)}$ are traceless tensors and $\langle.\,, . \rangle$ denotes the inner product contracted over $n$ dimensions \cite{MultipoleMethod}.
Specifically, contributions up to the quadrupole moment are calculated as,
\begin{equation}
    \begin{aligned}
        V_{M^{(0)}} &= \frac{1}{N}\sum_j^N\sum_i \frac{M_i^{(0)}}{|R_{ij}|}\\
        V_{M^{(1)}} &= V_{M^{(0)}} + \frac{1}{N}\sum_j^N\sum_i \frac{M_{i, \alpha}^{(1)}R_{ij, \alpha}}{|R_{ij}|^3}\\
        V_{M^{(2)}} &= V_{M^{(1)}} + \frac{1}{N}\sum_j^N\sum_i \frac{M_{i, \alpha\beta}^{(2)}(3R_{ij, \alpha}R_{ij, \beta} - R^2\delta_{\alpha\beta})}{2|R_{ij}|^5}\\
    \end{aligned}
\end{equation}
with $i$ iterating over multipole sites and $j$ iterating over a set of points at which the ESP is probed. The averaged ESP obtained in this manner is used to validate the ESP derived from multipoles.

\subsection{Loss Functions}
We define the mean squared error for the multipole moments of order $k$, which was used during training and validation, as
\begin{equation}\label{eq:mse}
    \mathcal{L}_2(M_{ref}^{(k)}, M_{pred}^{(k)}) = \frac{1}{AN}\sum_i^N\sum_\alpha^{A} (M^{(k)}_{i\alpha, \text{ref}} - M^{(k)}_{i\alpha, \text{pred}})^{2} ,
\end{equation}
with $N$ running over every atom of the molecule and $\alpha$ over each unique component with $A=\frac{(k+1)(k+2)}{2}$. 
The mean absolute error (MAE) used for model validation is defined analogously by replacing the $\mathcal{L}_2$ with the $\mathcal{L}_1$ norm.

\subsection{Dataset}
A dataset of multipoles was constructed for a set of molecules taken from three sources. First, neutral molecules with up to $20$ heavy atoms consisting of elements $\in \{\text{H, C, N, O, F, S, Cl}\}$ were selected from the ChEMBL \cite{CHEMBL11, CHEBML13, CHEMBL16} database. 
To improve coverage of very small molecules, systems from GDB11\cite{GDB} with up to eight heavy atoms and systems from a recently published dataset of dimer dissociation curves\cite{DEShawDimers} were added.
Due to their uniqueness, molecules from the GDB11 and dimer datasets with up to four heavy atoms (135 in total) were exclusively assigned to the training set.
All the other molecules were randomly and exclusively assigned to the training, validation, or test set. The final training set consisted of $272'924$ molecules, the validation set of $32'755$ molecules, and the test set contained $993$ molecules.

The ETKDG conformer generator,\cite{ETKDG} version 2 as implemented in the RDKit\cite{RDKIT} was used to generate 3D conformations.
For the molecules from the GDB11 and dimer datasets, up to four conformations were generated using an RMS pruning threshold of $0.1$~\AA.
For the very small systems (up to four heavy atoms), up to $16$ additional conformations were generated by randomly perturbing the generated conformation with Gaussian noise ($\sigma = 0.05$~\AA).
For all other compounds, up to two conformations were generated for the molecules in the training set and validation set, and up to $16$ for the test set using a RMS pruning threshold of $0.5$~\AA.
A larger number of conformations was produced for the test set to evaluate conformational dependence, resulting in a total of $737'280$ conformations for the training set, $117'312$ for the validation set, and $13'344$ for the test set. 

In addition to the fixed split, a five-fold cross validation was performed by randomly selecting one conformation for each unique molecule resulting in a total of $307'200$ data points. 
The data points were split into five folds. For each validation cycle, one fold was used for validation while the remaining folds were used to train the model.

DFT single-point calculations were performed using the PSI4 (version 1.4) package \cite{PSI4, PSI4_1.1} on a PBE0-D3BJ/def2-TZVP level of theory \cite{PBE, PBE0, DEF2, D3, D3BJ, DFT1, DFT2}. If not noted otherwise, default PSI4 settings were used. While not having any effect on the multipoles derived from single-point calculations, the D3BJ dispersion correction was included for future use of gradients and potential energies.
Distributed multipoles up to the octupole were extracted for all converged DFT single-point calculations with MBIS \cite{MBIS} as implemented in the recent PSI4 version \cite{PSI4_1.4}. Calculations were performed on the Euler HPC cluster at ETH Z\"urich.

\subsection{Model Parameters and Training}
$h_i^0$ was initialized with a one-hot encoding of the respective element type while each edge $a_{ij}$ contained the kernel-expanded distance between $i$ and $j$ using evenly spaced triweight kernels,
\begin{equation}
    f(x)=\text{max}(0, k(m-x^2)^3) ,
\end{equation}
with a shape factor $k$ and the center $m$. $32$ evenly spaced kernels covering the range ($0.5\,$\AA, $4\,$\AA) were used. The triweight kernel was chosen because its function value and first derivative take on $0$ at the cutoff.
Edges were added for all atom pairs within a cutoff radius of $4.0$~\AA. No strong dependence on the cutoff distance was observed when testing models based on cutoffs between $3$~\AA and $5$~\AA (data not shown). 
Molecular topology-based graphs were constructed by adding an edge for each chemical bond. Node features include the one-hot encoded element type as well as the scalar encoded hybridization state. Edges features include the two onehot encoded element types of the bonded atoms and four scalars encoding the bond type, aromaticity, conjugacy and ring membership. RDKit was used to obtain this information \cite{RDKIT}. All other parameters were kept consistent with geometry-based graphs.
Node- and edge-feature dimensions were set to $128$. Each GNN layer consisted of two fully feed-forward layers with $128$ units each ($[128, \text{Mila}, 128, \text{Mila}]$) combined with the Mila nonlinearity with $\beta=-1$ \cite{Mila}. The same architecture was used for the output layers $\phi_x$ ($[128, \text{Mila}, 128, \text{Mila}, 1, \text{Linear}]$). Layer weights were initialized with the method introduced by He \cite{HE}. The graph was updated over four message passing iterations. 
Further, a residual architecture was used for atom feature updates \cite{ResNet},
\begin{equation}
    h_i^{l+1} = h_i^l + \phi_h(h_i^l, m_i) .
\end{equation}

The model was trained over $2'048$ epochs, presenting $2'048$ randomly drawn batches of size $64$ during each epoch. 
Weights were updated using the ADAM optimizer \cite{ADAM} with an exponentially decaying learning rate $[5\cdot 10^{-4}, 10^{-5}]$, minimizing the mean-squared error between reference multipole components and predicted multipole components as defined in Eq.~\ref{eq:mse}. Models for monopoles, dipoles, and quadrupoles were optimized independently. Shared weight models were found to provide slightly worse results at lower computational costs. 
For the five-fold cross-validation, a model was randomly initialized for each fold and trained over $512$ epochs with $2'048$ randomly drawn batches of size $16$. All other parameters remained unchanged.
Batch-size and number of epochs were decreased to accelerate the training, resulting in a slightly lower accuracy for the cross-validation models compared to the main model. 

The models were implemented with TensorFlow (2.6.2) \cite{TensorflowPaper, TF2.6.2} and the GraphNets library (1.1.0) \cite{GNN} using the InteractionNetwork model \cite{InteractionNetwork}. 
General code was written with Python (3.9.7) \cite{Python} and Numpy (1.19.5) \cite{Numpy}.
Plots and visualizations were made with Open3D (0.13.0) \cite{Open3D}, Matplotlib (3.4.2) \cite{PLT}, Seaborn (0.11.0) \cite{Seaborn}, and VMD (1.9.3) \cite{VMD}. 

\section{Results and Discussion}
The results are structured in two parts: First, the prediction accuracy of the ML model is assessed for the explicitly trained quantity, i.e. the multipole components. Second, properties derived from predicted multipoles are used to validate the model. In particular, the ESP reconstructed from predicted multipoles is compared with the reference ESP obtained from DFT or reconstructed from reference multipoles.

\subsection{Prediction Accuracy on Multipoles}
Table~\ref{tab:mae} lists the mean absolute errors (MAE) of the predicted multipoles with respect to the reference multipoles for the specified subsets. Errors are given in $10^{-3}$ elementary charge. The results indicate that the proposed model is able to predict multipoles up to the quadrupole with high accuracy. While a direct comparison is difficult due to different dataset sizes, difficulties and employed reference methods, there exists, to our knowledge, currently no model in the literature which is able to predict atomic multipoles up to the quadrupole with better or comparable accuracy.

\begin{table*}[htb]
    \centering\begin{tabular}{@{}rrrrcrrrcrrr@{}}
    \hline
    \multicolumn{3}{c}{MAE $M^{(0)}$ [me]} & \phantom{abc} & \multicolumn{3}{c}{MAE $M^{(1)}$ [me$\cdot$\AA]} & \phantom{abc} & \multicolumn{3}{c}{MAE $M^{(2)}$ [me$\cdot$\AA$^2$]}\\
    Train & Val & Test && Train & Val & Test && Train & Val & Test\\\hline
    2.05 & 2.20 & 2.19 && 
    0.64 & 0.65 & 0.64 && 
    0.63 & 0.64 & 0.63 \\ 
    \hline
    \end{tabular}\caption{Mean absolute errors for the predicted multipoles of the training, validation, and test sets: monopoles in $10^{-3}$ elementary charge [me], dipoles in [me$\cdot$\AA], and quadrupoles in [me$\cdot$\AA$^2$].}
    \label{tab:mae}
\end{table*}

\begin{figure}[H]
  \begin{subfigure}{\textwidth}
    \centering
    \includegraphics[width=\textwidth]{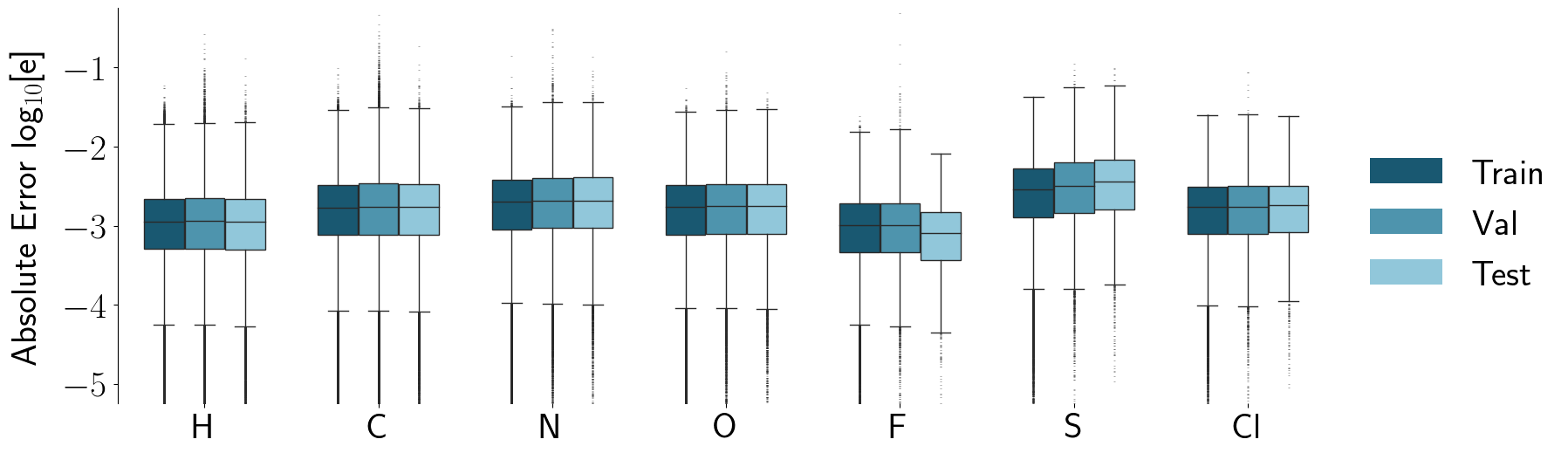}
    \caption{Absolute error of monopoles in log$_{10}$ [e].}
  \end{subfigure} \hfill
  \begin{subfigure}{\textwidth}
     \includegraphics[width=\textwidth]{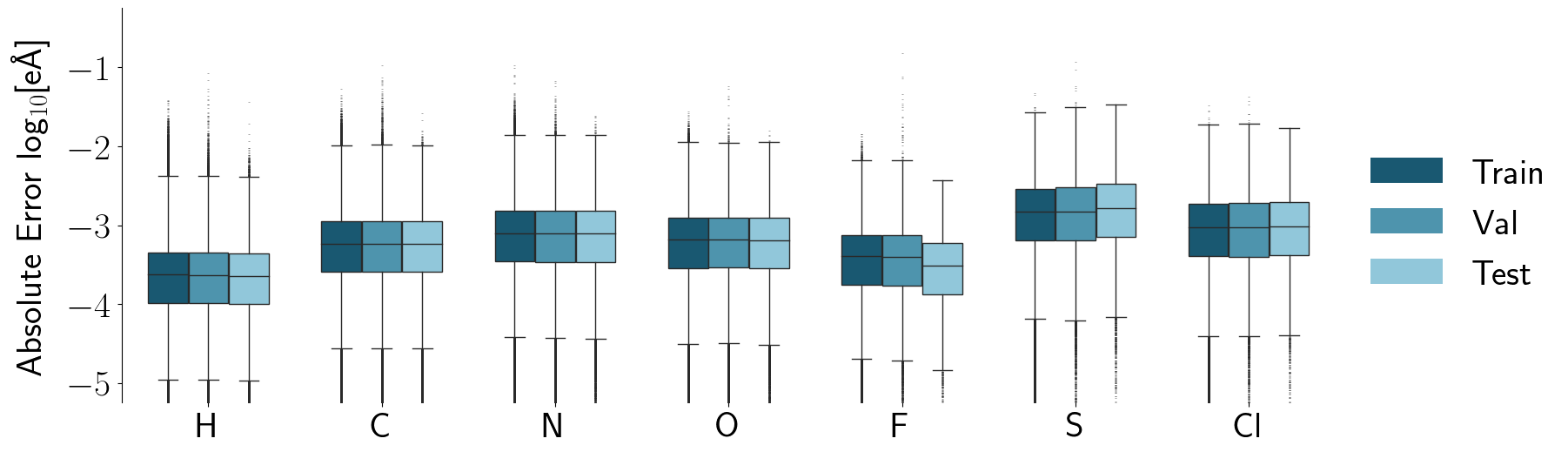}
     \caption{Absolute error of dipole components in log$_{10}$ [e$\cdot$\AA].}
  \end{subfigure} \hfill
  \begin{subfigure}{\textwidth}
     \includegraphics[width=\textwidth]{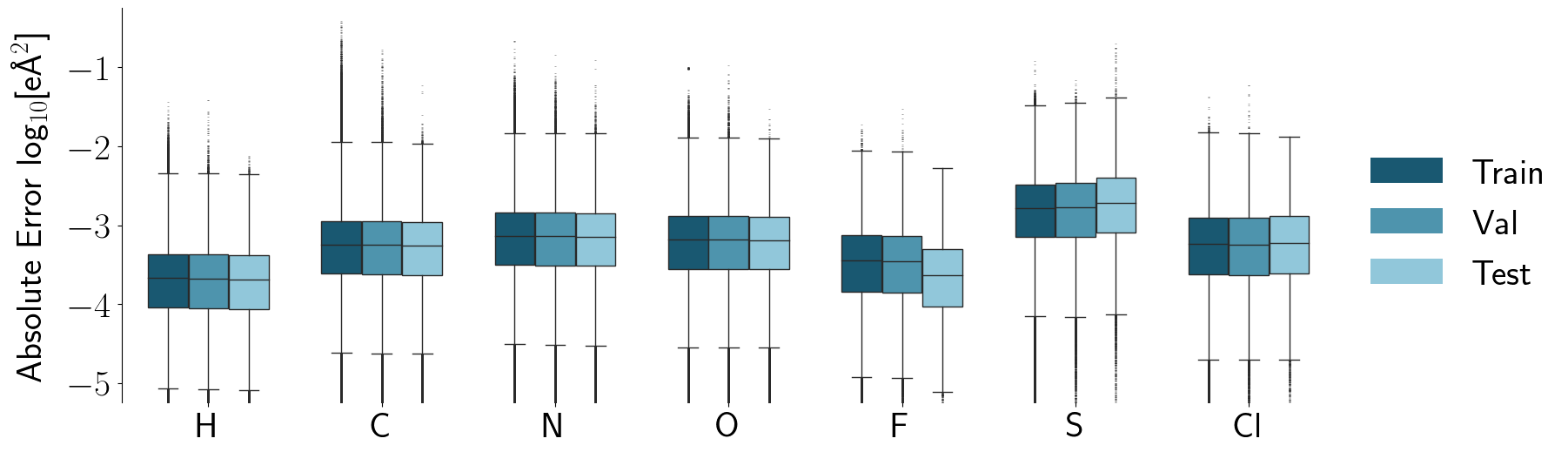}
     \caption{Absolute error of quadrupole components in log$_{10}$ [e$\cdot$\AA$^2$].}
  \end{subfigure}
    \caption{Distribution of the absolute error of the multipole components by element for the training, validation, and test sets.}
    \label{fig:elementtypes}
\end{figure}

When separated into element types, it is evident that the errors are not distributed equally for each element (Figure \ref{fig:elementtypes}). While the frequency of each element in the respective data set (Table~S2 in the Supporting Information) 
plays a role, the largest factor seems to be the range of the respective reference multipole components. 
This effect can be seen most clearly for sulphur whose monopoles cover a range of more than two elementary charges
(Figures S1 - S3 in the Supporting Information) and for which element the observed error is largest. 
On the other hand, the monopoles of fluorine atoms cover only a fraction of this range resulting in the smallest observed error.
In addition, molecule size is a contributing factor. Particularly for very small molecules, which are unique by nature and less abundant in the dataset, larger error fluctuations can be observed (Figures S4 - S10 in the Supporting Information).

The robustness of the model performance and its ability to generalize was assessed with a five-fold cross validation. For this purpose, one conformation was randomly sampled for each unique molecule such that each molecule was exclusively present in the training or the validation set, but not both. These selected data points were randomly assigned to one of five equal sized tranches resulting in $61'440$ data points for each tranche, and a total of $307'200$ samples.
For each fold, one tranche was used as the validation dataset while the other four tranches were used to train the model.
The average MAE and standard deviation over the five folds is listed in Table \ref{tab:crossvalidation}. The small standard deviations observed indicate a robust performance of the model. Note that models were trained less extensively for the cross-validation compared to the main model, resulting in slightly larger errors.

\begin{table*}[htb]
    \centering\begin{tabular}{@{}ccc@{}}
    \hline
    \multicolumn{3}{c}{MAE Cross Validation}\\
    \phantom{abc} & Train (4 folds) & Val (1 fold) \\\hline
    $M^{(0)}$ & 2.684 $\pm$ 0.017 & 2.918 $\pm$ 0.014 \\
    $M^{(1)}$ & 0.848 $\pm$ 0.005 & 0.863 $\pm$ 0.003 \\
    $M^{(2)}$ & 0.826 $\pm$ 0.008 & 0.845 $\pm$ 0.008 \\
    \hline
    \end{tabular}
    \caption{Mean absolute errors (MAE) for the predicted multipoles from a five-fold cross validation. Mean and standard deviation are taken over the MAE of the five folds.  The data set consisted of all data points in the training and validation sets from the fixed split: monopoles is given in $10^{-3}$ elementary charge [me], dipoles in [me$\cdot$\AA], and quadrupoles in [me$\cdot$\AA$^2$].}
    \label{tab:crossvalidation}
\end{table*}

\subsection{Accuracy of the Electrostatic Potential}
To validate the fidelity with which multipoles reproduce the ESP, electron densities from PBE0-D3BJ/def2-TZVP calculations were used to obtain the ESP ($\text{ESP}_{\text{ref}}$) on a surface with a distance of two vdW radii around the molecule. For this case, up to $16$ conformations were generated for each molecule in the test set. On average, each surface contained around $4'000$ points. For the calculation of the ESP from multipoles, a probe charge of $1\,$e was placed on each point on the vdW surface.

Table~\ref{tab:esp_mae} lists the MAE with respect to the reference potential $\text{ESP}_{\text{ref}}$ for systems in the test set.
The errors are averaged over $13'344$ conformations of $997$ unique molecules from the test set. Three molecules had to be excluded due to convergence issues.
In addition to the analytically derived reference multipoles, which serve as the baseline, predicted multipoles and partial charges are compared. Further, truncations after monopoles, dipoles and quadrupoles ($M^{(0)}$, $M^{(1)}$, $M^{(2)}$) are provided.
As a comparison we also provide errors obtained from a model which uses molecular graphs without distance information as input. 
These entries are labeled with the subscript ``topo'' in Table \ref{tab:esp_mae}. The prediction accuracy of this model is provided in Table S1 in the Supporting Information. 
In order to better understand the error composition, the error obtained by only using predicted multipoles for the highest considered moment is listed using a $\Delta$ symbol, i.e. $\Delta M^{(1)}$ denotes the error of the ESP based on reference monopoles and predicted dipoles, while $\Delta M^{(2)}$ denotes the error obtained by using reference monopoles and dipoles in combination with predicted quadrupoles.

\begin{table*}[htb]
    \centering\begin{tabular}{@{}ccc|ccc|cc|ccc@{}}
        \hline
        \multicolumn{11}{c}{MAE $V_{\text{ES}}$ [kJ$\cdot$mol$^{-1}$]}\\
        $M^{(0)}_{\text{ref}}$ & $M^{(1)}_{\text{ref}}$ & $M^{(2)}_{\text{ref}}$ & $M^{(0)}_{\text{pred}}$ & $M^{(1)}_{\text{pred}}$ & $M^{(2)}_{\text{pred}}$ & 
        $\Delta M^{(1)}$ &
        $\Delta M^{(2)}$ &
        $M^{(0)}_{\text{topo}}$ & 
        $M^{(1)}_{\text{topo}}$ & 
        $M^{(2)}_{\text{topo}}$ 
        \\\hline
        4.23 & 2.74 & 0.77 & 
        4.38 & 2.95 & 1.30 &  
        2.76 & 0.80 &  
        6.58 & 6.03 & 5.32 \\ 
        \hline
    \end{tabular}\caption{Mean absolute errors (MAE) of the ESP on the vdW surface in kJ$\cdot$mol$^{-1}$ based on reference and predicted multipoles with respect to the DFT ESP. 
    Absolute errors are averaged over $13'344$ data points from the test set. The order $k$ of $M^{(k)}$ refers to the highest multipole term considered. Subscript ``pred'' refers to predicted multipoles, while ``ref'' denotes the reference multipoles based on the DFT density. For the entries labeled with $\Delta$, only the highest moment was predicted by the model while lower multipoles were analytically derived.
    Entries denoted by ``topo'' were obtained from a model, which uses only molecular graphs as input without distance information.}
    \label{tab:esp_mae}
\end{table*}

When using the analytically derived reference multipoles, a clear reduction of the MAE can be observed when including dipoles ($M^{(1)}_{\text{ref}}$) and quadrupoles ($M^{(2)}_{\text{ref}}$). The same trend is seen for the predicted multipoles.
The results shown in Table \ref{tab:esp_mae} show an accuracy hierarchy and accumulation of the error when including higher multipoles. In absolute terms, ignoring the conformational dependence of monopoles (entries labeled with ``topo'') results clearly in the largest error, followed by an early truncation of the multipole series. Replacing analytically derived multipoles with ML predicted multipoles introduces an error of $0.15$, $0.21$ and $0.53$ kJ$\cdot$mol$^{-1}$, respectively. $\Delta M^{(1)}$ and $\Delta M^{(2)}$ indicate that the error introduced by each predicted component itself is comparatively small, however, errors accumulate with each term of the multipole series. In cases where the ESP is approximated using the interaction between more than one multipole source, the error might be amplified additionally since each moment interacts with all other moments. This is not the case in the validation presented here where each multipole source interacts only with a monopole probe charge. Nevertheless, the results also clearly indicate that the error incurred by the ML model is in all cases smaller than the error caused by a premature truncation. These results suggest that the use of ML predicted multipoles up to quadrupoles in the treatment of electrostatic interactions is preferable over calculations with only analytically derived monopoles (or dipoles).

\begin{figure}[H]
    \centering
    \includegraphics[width=\textwidth]{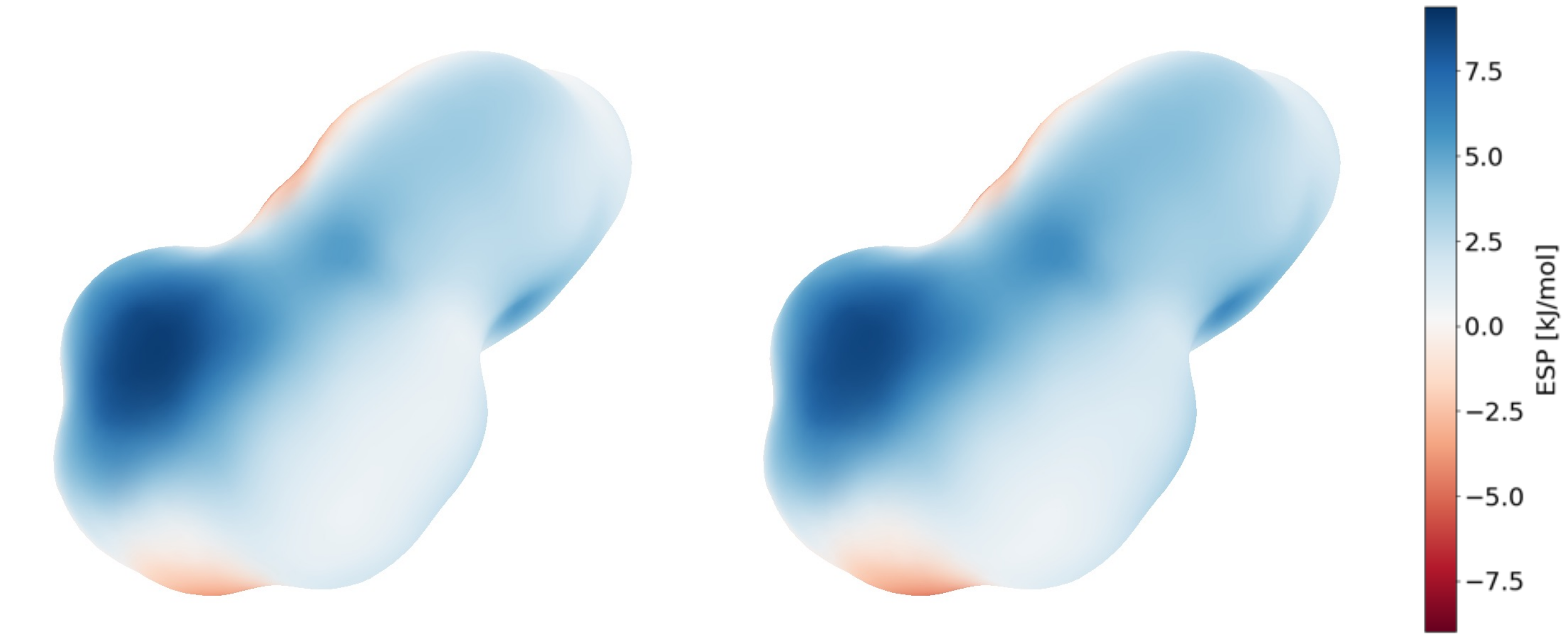}
    \caption{Delta ESP of the ESP based on analytically derived reference monopoles (left, $V_{\text{ES, DFT}}-V_{\text{ES, $M^{(0)}_\text{ref}$}}$) and predicted monopoles (right, $V_{\text{ES, DFT}}-V_{\text{ES, $M^{(0)}_\text{pred}$}}$) with respect to the DFT ESP.}
        \label{fig:delta_esps}
\end{figure}

\begin{figure}[H]
    \centering
    \includegraphics[width=\textwidth]{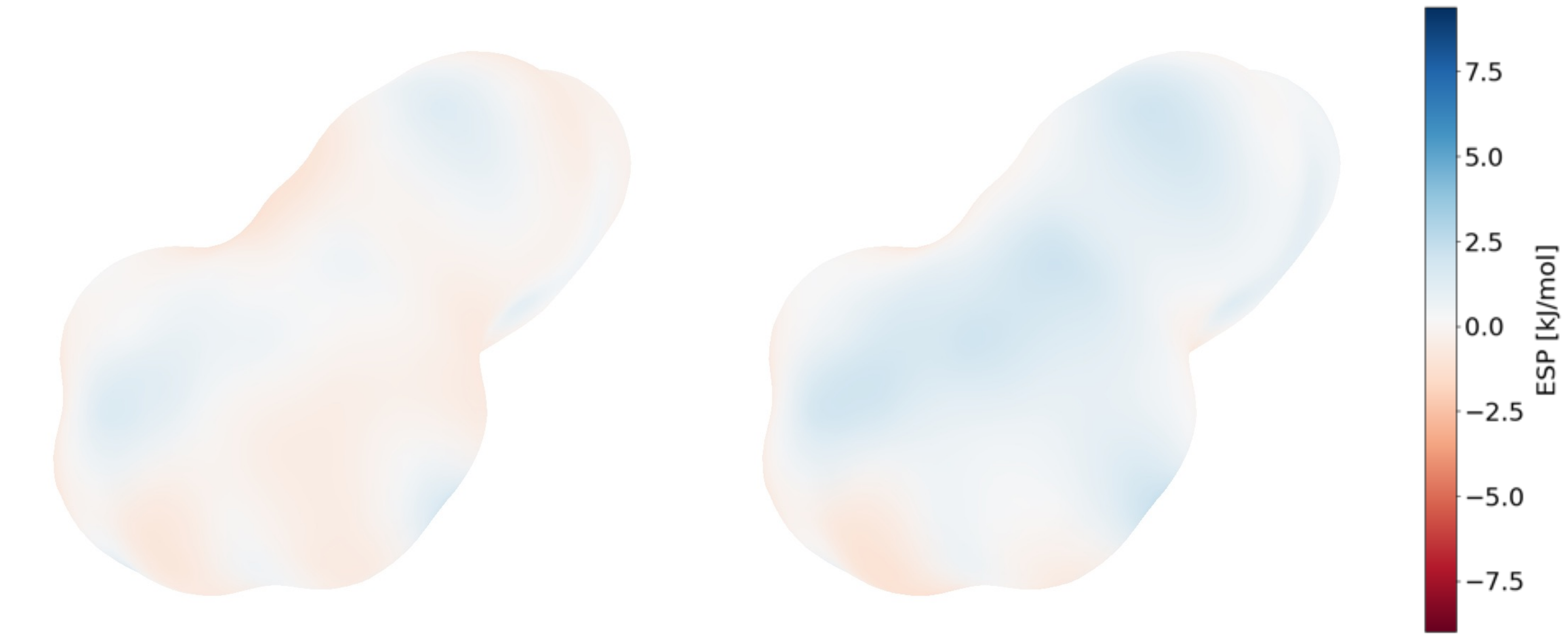}
    \caption{Delta ESP of the ESP based on analytically derived reference multipoles up to the quadrupole (left, $V_{\text{ES, DFT}}-V_{\text{ES, $M^{(2)}_\text{ref}$}}$) and predicted multipoles up to the quadrupole (right, $V_{\text{ES, DFT}}-V_{\text{ES, $M^{(2)}_\text{pred}$}}$) with respect to the DFT ESP. The same scale as in Figure \ref{fig:delta_esps} is used.}
    \label{fig:delta_esps_multi}
\end{figure}

To visually illustrate the results, the DFT ESP and predicted ESP on the vdW surface are compared for an example molecule (5-((4-Fluorobenzyl)oxy)benzofuran-3(2H)-one) randomly selected from the test set. It can be clearly seen in Figure \ref{fig:delta_esps} that monopoles are insufficient to reproduce the more intricate features of the ESP. Including dipoles and quadrupoles reduces the error substantially (Figure \ref{fig:delta_esps_multi}). While only one illustrative case, this example shows that even though predicted multipoles result in a slightly less accurate ESP compared to analytically derived multipoles, predicted multipoles correctly reproduce qualitative features of the ESP and improve the accuracy (Figure \ref{fig:delta_esps_multi}).

\begin{figure}[H]
    \centering
    \includegraphics[width=\textwidth]{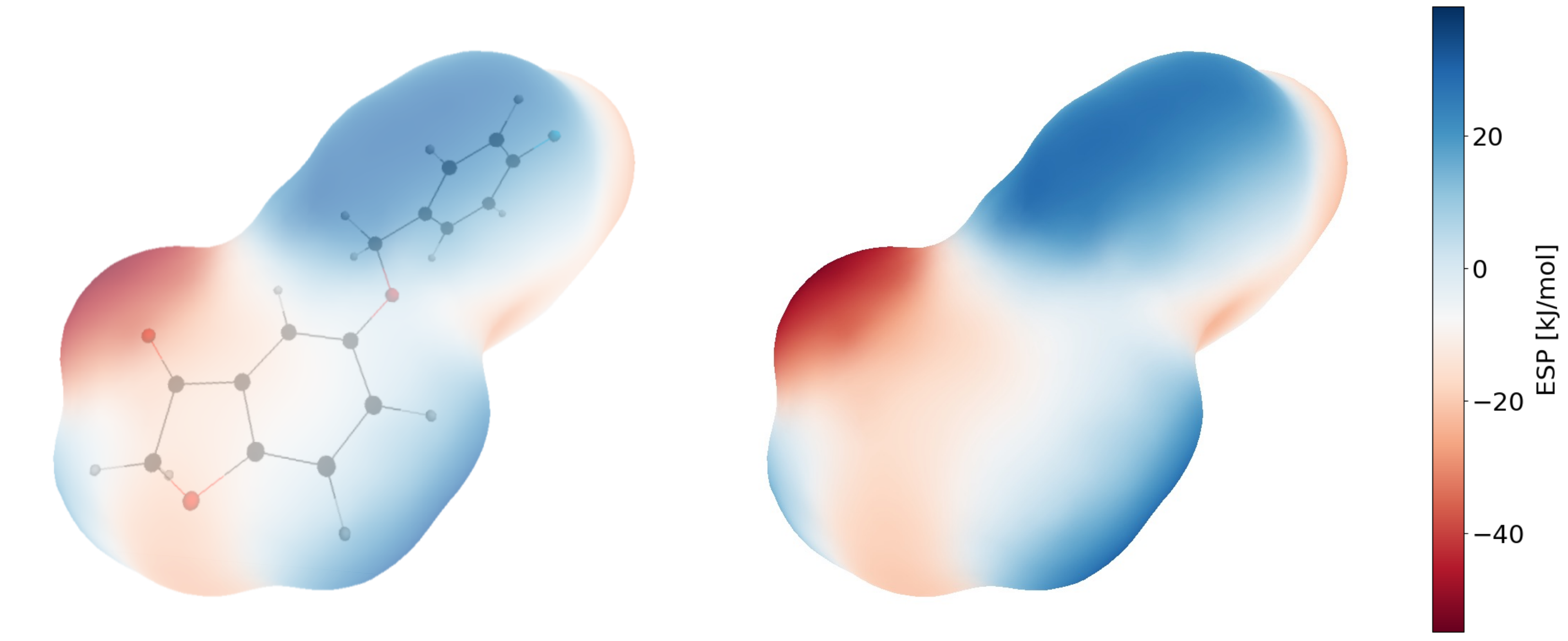}
    \caption{DFT ESP (left, $V_{\text{ES, DFT}}$) and ESP derived from predicted multipoles (right, $V_{\text{ES, $M^{(2)}_\text{pred}$}}$) using the same scale with different transparencies.}
    \label{fig:esps}
\end{figure}

For derived properties, such as the ESP or the molecular dipole, error cancellation when including higher-order terms is desirable. For specific future applications, this might be achieved by introducing additional loss terms. Particularly, a loss term for the molecular dipole could be introduced based on the error between a reference molecular dipole and a predicted molecular dipole constructed from predicted atomic monopoles and predicted atomic dipoles. In addition, a term for the error between the ESP based on predicted multipoles and reference multipoles could be added by randomly sampling the ESP around a given molecule. 

\subsection{Timing}
Model speed was tested on a NVIDIA Titan V using the same batch with $50'064$ atoms for all models (Table \ref{tab:timing}). In all cases, single precision was used. Single precision did not lower the model accuracy but resulted in a two-fold speed-up on a GPU compared to double precision.
The execution time was averaged over $100$ runs with four repeats each, and are given in $\mu$s$/$atom.
To explore the dependence on model complexity and graph size, we also included a model that uses shared layers (for the embedding step and a shared layers for the dipole and quadrupole coefficients, i.e. one module with two outputs). The same model parameters and model architecture were used for all models, consistent with the main model used for the results reported in the previous section. The time needed to construct the graph is not included as it can be sped up drastically in repeated calculations. 
Without specifically taking advantage of redundant calculations, the construction of a graph takes around $0.256\,\mu$s/atom for a molecule with $25$ atoms. At $0.075\,\mu$s/atom and $0.026\,\mu$s/atom, the construction of the edge features as well as distance calculations and pair extraction are the only steps, which cannot be circumvented in repeated calculations.
In addition, execution speed can be improved further by reducing the number of graph-updating steps $n$ or the cutoff radius $r_c$ without strongly hampering prediction accuracy.
\begin{table*}[htb]
    \centering\begin{tabular}{@{}ccc@{}}
    \hline
    \multicolumn{3}{c}{Execution Time (Prediction) [$\mu$s/atom]}\\
    $n$ & Main Model & Shared Layers \\\hline
    3 & 6.53 $\pm$ 0.04 & 3.26 $\pm$ 0.08 \\\hline
    4 & 8.21 $\pm$ 0.05 & 3.86 $\pm$ 0.08\\\hline
    \end{tabular}
    \caption{Time required to predict multipoles for a single atom given a pre-constructed graph. Numbers in brackets refer to the cutoff radii used for the graph construction (r$_c$). $n$ refers to the number of graph-updating steps used.}
    \label{tab:timing}
\end{table*}

\section{Conclusion}
In the present work, we have introduced an equivariant GNN for the prediction of atomic multipoles. The proposed model is able to reproduce the ESP on a hybrid-DFT level of theory with an accuracy close to that of analytically derived multipoles. The uncertainty introduced by the ML prediction is consistently smaller than the error caused by the omission of higher-order terms (i.e. dipoles, quadrupoles). Thus, compared to calculations involving only analytically derived monopoles, predicted multipoles enable a much more precise treatment of the electrostatic interaction at a fraction of the cost of QM reference calculations. The equivariant architecture used in the GNN is not only able to predict multipole components with high precision, it also strictly enforces the proper behaviour under rotations of the system without relying on local reference frames.
Further, the sparse representation, which constructs atomic multipoles as a weighted combination of vectors and tensor products of vectors to an atom's neighbours, performs well for unseen structures in the test set.

Multipoles predicted by the proposed model may be used to describe the static multipoles in polarizable FFs, the analysis of ESP surfaces, or for the treatment of long-range interactions in ML models. 
For future work, handling of charged species, extended coverage of the chemical and conformational space as well as including an explicit treatment of polarization from the environment, which is important for condensed-phase systems, could be of value. In addition, directional message passing or various other atomic environment descriptors, may further improve accuracy.

\section*{Data and Software Availability}
The full dataset covering $311'781$ unique molecules and $1'013'949$ conformations on a PBE0-D3BJ/def2-TZVP level of theory, including potential energies, gradients, MBIS multipoles up to the octupole, as well as ESP surfaces for the test set, is publicly available on the ETH Research Collection (\url{https://www.research-collection.ethz.ch/handle/20.500.11850/509052}). A repository with the ML model used to generate the results in this work can be found at \url{https://github.com/rinikerlab/EquivariantMultipoleGNN}.

\section*{Acknowledgements}
M.T. was supported by the NCCR MARVEL, funded by the Swiss National Science Foundation. The authors thank NVIDIA for providing a Titan V under its Academic Hardware Grant Program.

\bibliographystyle{b2}
\bibliography{references}

\end{document}